\documentclass[a4paper,11pt]{article}
\pdfoutput=1
\usepackage[utf8]{inputenc}
\usepackage{jinstpub}
\usepackage{ulem}
\usepackage{multirow}
\usepackage{float}

\title{AI for Experimental Controls at Jefferson Lab}
\author[a]{Torri Jeske,}
\author[a]{Diana McSpadden,}
\author[a]{Nikhil Kalra,}
\author[a]{Thomas Britton,}
\author[b]{Naomi Jarvis,}
\author[a]{and David Lawrence}

\affiliation[a]{Thomas Jefferson National Accelerator Facility, Newport News, VA 23606, USA}
\affiliation[b]{Department of Physics, Carnegie Mellon University, Pittsburgh, PA 15213, USA}

\emailAdd{roark@jlab.org}
\emailAdd{davidl@jlab.org}
\date{November 2021}

\abstract{The AI for Experimental Controls project is developing an AI system to control and calibrate detector systems located at Jefferson Laboratory. Currently, calibrations are performed offline and require significant time and attention from experts. This work would reduce the amount of data and the amount of time spent calibrating in an offline setting. The first use case involves the Central Drift Chamber (CDC) located inside the GlueX spectrometer in Hall D. We use a combination of environmental and experimental data, such as atmospheric pressure, gas temperature, and the flux of incident particles as inputs to a Sequential Neural Network (NN) to recommend a high voltage setting and the corresponding calibration constants in order to maintain consistent gain and optimal resolution throughout the experiment. Utilizing AI in this manner represents an initial shift from offline calibration towards near real time calibrations performed at Jefferson Laboratory. }

\begin{document}

\maketitle

\section{Introduction}
\begin{figure}[htpb]
    \centering
    \includegraphics[width=.7\textwidth]{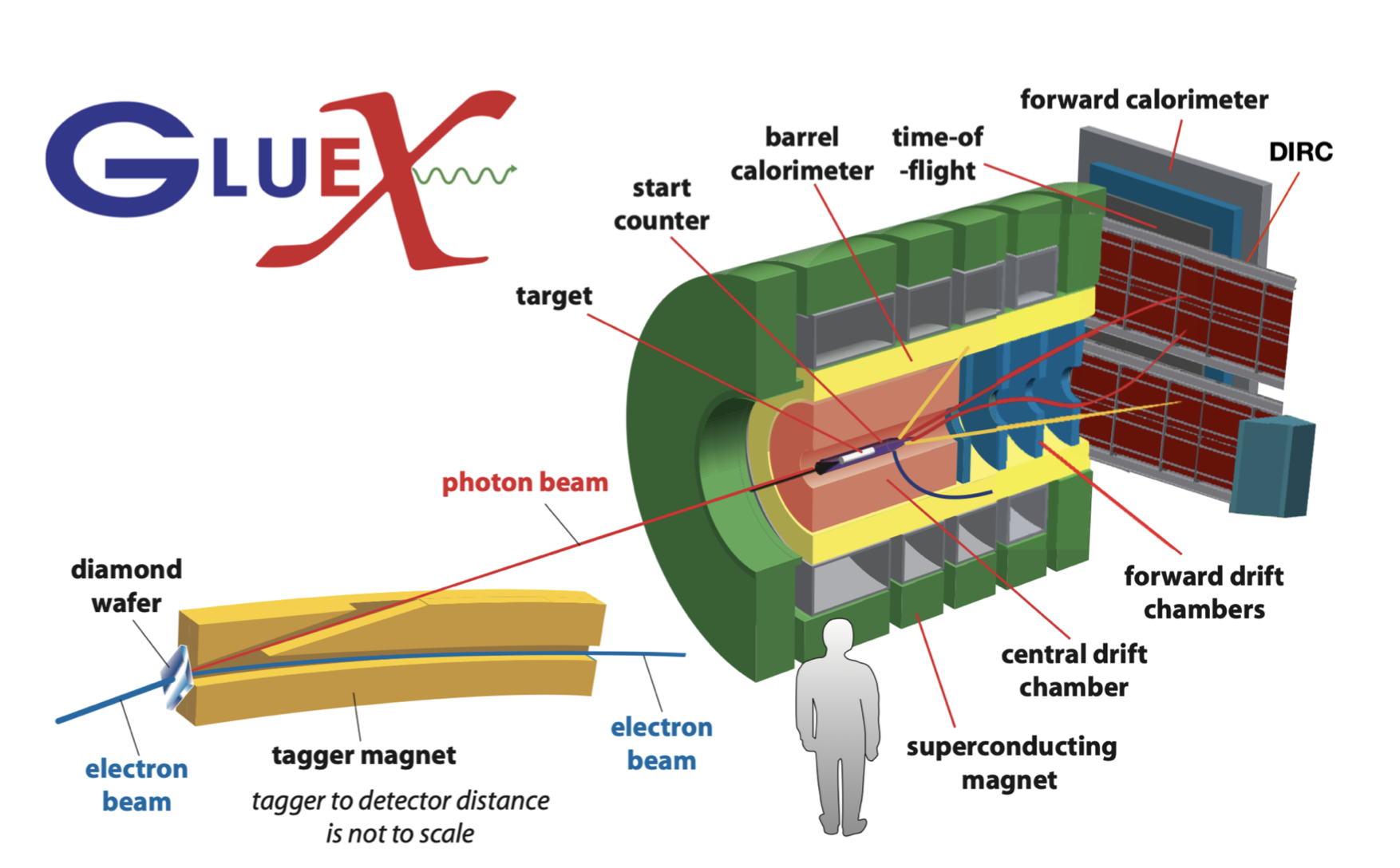}
    \caption{Schematic of the GlueX Spectrometer located in Hall D at Jefferson Laboratory.}
    \label{fig:gluex}
\end{figure}

The Gluonic Excitations (GlueX) experiment was designed to reconstruct exclusive final states from photoproduction reactions on proton targets using the GlueX Spectrometer located in Hall D at Jefferson Laboratory \space \cite{GlueX_Detector2021}. The GlueX collaboration consists of approximately 150 members across 27 institutions. The primary objective of the GlueX experiment is to search for and measure exotic mesons predicted from Lattice QCD. A schematic of the GlueX spectrometer is shown in Figure \ref{fig:gluex}.

\begin{figure}[htpb]
    \centering
    \includegraphics[width=0.8\textwidth]{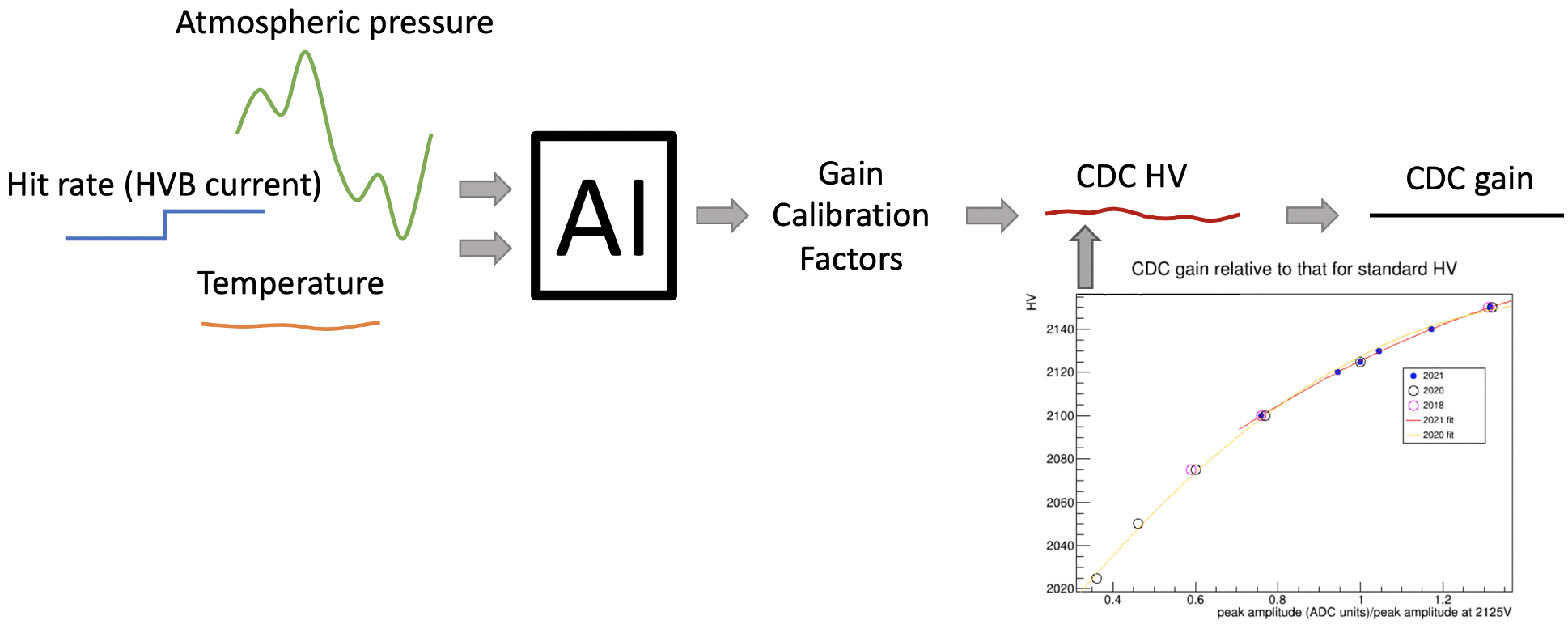}
    \caption{Illustration of the AI model. The input features consists of environmental and experimental data and the outputs are the HV settings for the next run and the associated calibration constants for the Central Drift Chamber located in GlueX.}
    \label{fig:AI_model}
\end{figure}

A typical experiment in GlueX takes data in "runs". The duration of each run is limited to two hours in large part to minimize changes in the drift chamber gain caused by fluctuations in the atmospheric pressure. The current calibration procedure for GlueX requires most calibrations to be performed \textit{after} the experiment has been completed. In the first round of calibrations, each detector is calibrated independently. In each subsequent round, information is shared between the detectors to refine the timing offsets and track information. Calibration constants are produced for every run taken during the run period. The time scale to complete all calibrations is on the order of months. In contrast, the AI system would be used \textit{during} the experiment. Before starting a new run, the input features are checked to ensure they are not drastically different from values used during model training. It would be given experimental and environmental data as input features, predict the calibration constants, and recommend the associated CDC high voltage (HV) setting for the next run in order to maintain consistent gain and resolution throughout the experiment. The recommendation for the HV setting is obtained from a fit to the chamber gain as a function of HV. An overview of this procedure is shown in Figure \ref{fig:AI_model}. As a first approach, the AI model is trained on the existing calibration constants from the 2018 and 2020 run periods. It is not anticipated that we would need to re-train any models for upcoming run periods unless there were significant changes in the experiment design or initial input features. 

\begin{figure}[htpb]
    \centering
    \includegraphics[width=.8\textwidth]{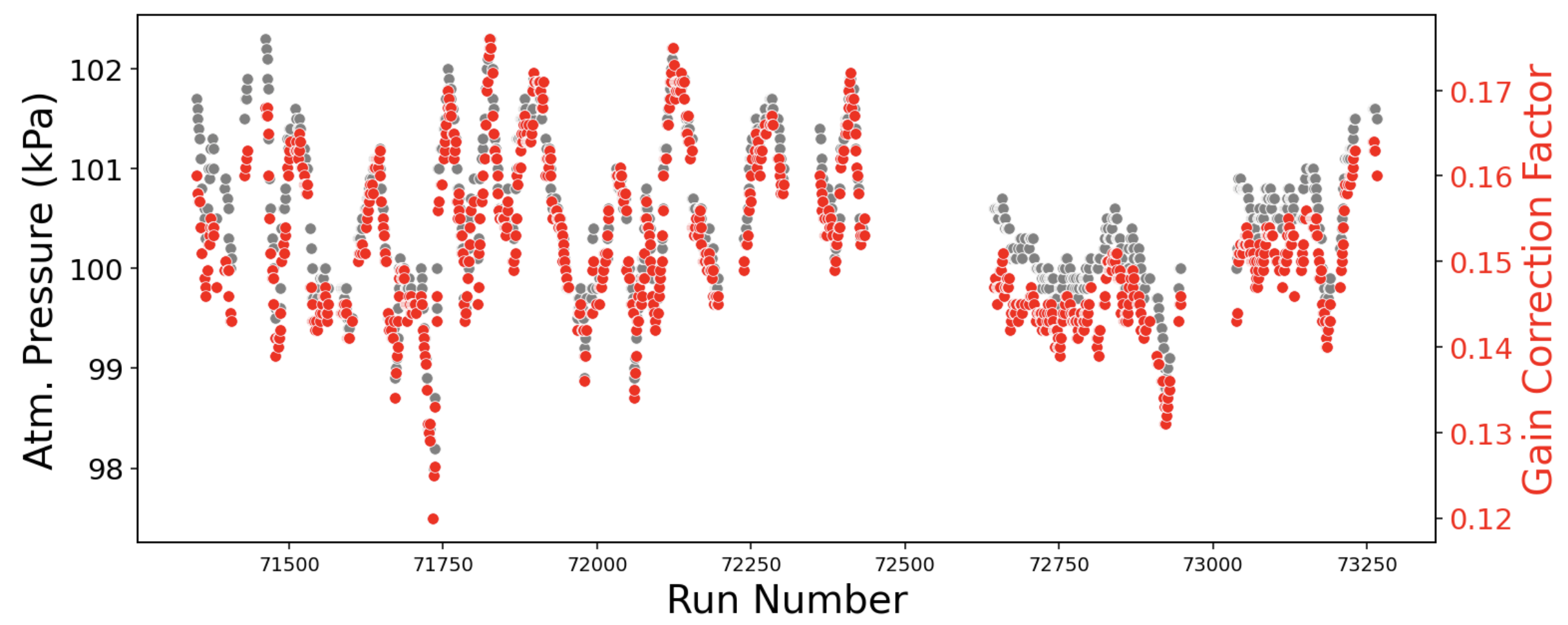}
    \caption{Atmospheric pressure (gray) and Gain Correction Factor (red) as a function of the run number during the Fall 2020 run period.}
    \label{fig:pressure_gain}
\end{figure}

The initial focus of this work is to control and calibrate the Central Drift Chamber used in GlueX \cite{CDC_NIMA}. The CDC is a cylindrical, straw tube drift chamber located inside of the GlueX solenoid. The primary purpose of the CDC is to detect and track charged particles as they pass through the detector. The chamber gain affects both the measured amplitude used for particle ID and the measured drift time used to determine the particle's momentum. There are a number of factors that affect the chamber gain, including: anode voltage, atmospheric pressure, gas temperature, hit rate, and beam current. The first calibration constant we aim to predict is the Gain Correction Factor (GCF). Conventionally, the GCF is obtained using the dE/dx of reconstructed tracks through the CDC. The GCF is a scaling factor applied to the CDC pulse height, and it is adjusted to
ensure that the experimental dE/dx matches that obtained from simulations. The GCFs obtained during the 2020 run period are shown in Figure \ref{fig:pressure_gain}. A single batch job per run is used to execute the tracking software needed to generate the required histograms for fitting. This process is computationally expensive and is performed after the experiment has ended. Another advantage of using the AI system is that it does not rely on any information from the tracking software.

\section{Model Development}
From the 2018 and 2020 run periods, 1,396 runs were available. The runs were divided into an 80/20 train test split, resulting in 1116 training observations and 280 testing observations. Training data include 345 runs from 2018 and 771 runs from the 2020 run period. Testing data includes 91 runs from 2018 and 189 runs from the 2020 run period. For each run, the beam current, atmospheric pressure, gas temperature, CDC high voltage board current, and radiation length was obtained from the Experimental Physics Industrial Controls (EPICS) system. From this data set, 122 features were derived, min-max scaled, and used for training. 

\begin{figure}[htbp]
    \centering
    \includegraphics[width=0.6\textwidth]{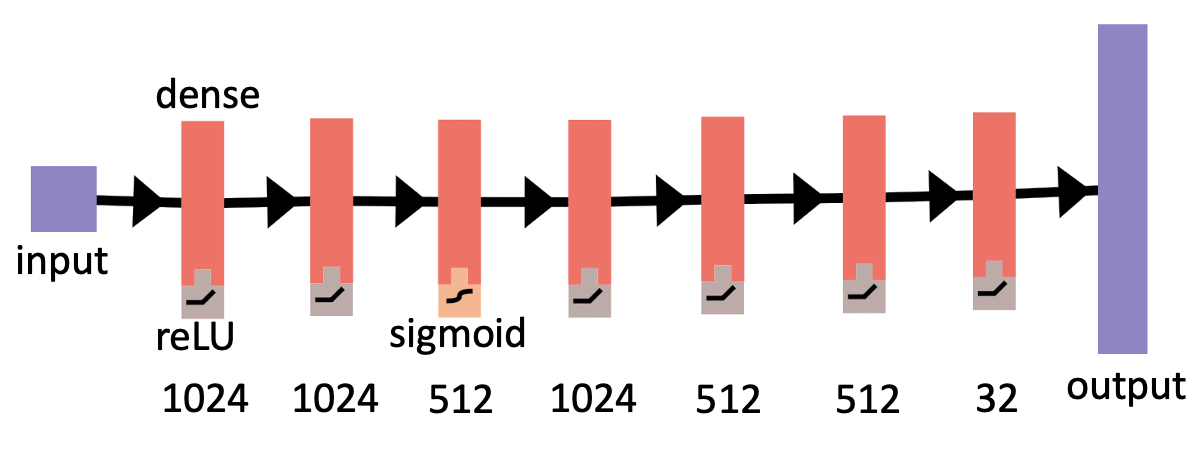}
    \caption{Sequential Neural Network architecture. The number of nodes and activation function is presented under each layer.}
    \label{fig:neural_net}
\end{figure}

The model architecture used for predicting the existing Gain Correction Factors (GCF) was selected using the Keras Tuner \cite{KerasTuner} package. Model architectures were generated from unique combinations in the hyper-parameter space of: 2-7 layers, learning rates of 0.001, 0.0005, 0.0002, 0.0001 and node widths of 16, 32, 64, 128, 512, 1024; and sigmoid and relu activation functions. The model with the minimum mean squared error is a multilayer perceptron consisting of 7 layers. The ADAM optimizer was used with a learning rate of 0.0002. The learning rate was decreased on plateaus during training. Training was stopped early if the loss remained stable for 50 epochs. The loss function is the percent error between the prediction and the existing GCF. The model architecture is shown in Figure \ref{fig:neural_net}. 

\begin{figure}[h]
    \centering
    \includegraphics[width=0.6\textwidth]{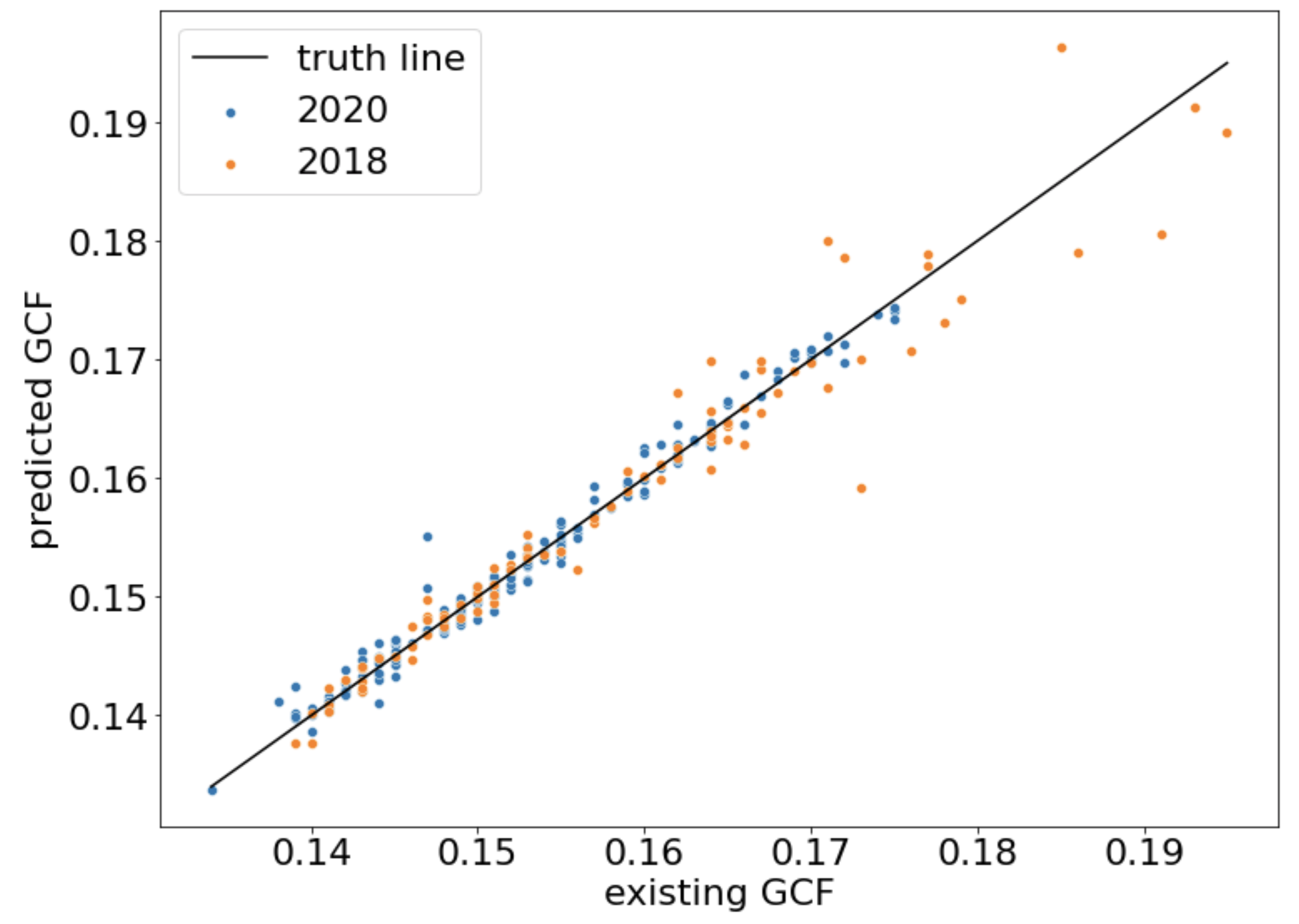}
    \caption{AI predicted GCFs compared to the existing GCFs for the 2018 (orange) and 2020 (blue) data sets. The solid black line is the truth line.}
    \label{fig:early_results}
\end{figure}
 
Figure \ref{fig:early_results} shows preliminary results of the predicted GCFs compared to the existing GCFs for both the Fall 2018 and 2020 data sets.

\begin{table}[H]
\centering
\begin{tabular}{ |p{3cm}||p{3cm}|p{3cm}|p{3cm}|  }
 \hline
 \multicolumn{4}{|c|}{NN Performance for 2018 and 2020 Run Periods} \\
 \hline
 & 2018 and 2020 Train and Test & 2018 Train and Test & 2020 Train and Test\\
 \hline
 Mean Percent Error   & 0.8\%    &1.3\%&   0.58\%\\
 Max Percent Error &   8.7\%  & 20.1\%   & 5.5\% \\
 Runs {$>$} 1\% Error & 57 of 280 & 31 of 91 & 33 of 189\\
 \hline
\end{tabular}
\caption{NN Performance by Run Period}
\label{nn_performance}
\end{table}

Table \ref{nn_performance} exhibits the increase in accuracy for the 2020 run period compared to the 2018 run period. It is known that the 2018 run period had a smaller data set and experienced a greater number of beam trips; however this doesn't entirely account for the decreased accuracy. We continue to explore the impact of a degrading diamond radiator, different ranges of pressure and temperature, varying beam current, and the time used to query the EPICS system compared to when the data from the run was recorded. 

\section{CDC Performance using AI}
\begin{figure}[H]
\centering
\includegraphics[width=.4\textwidth]{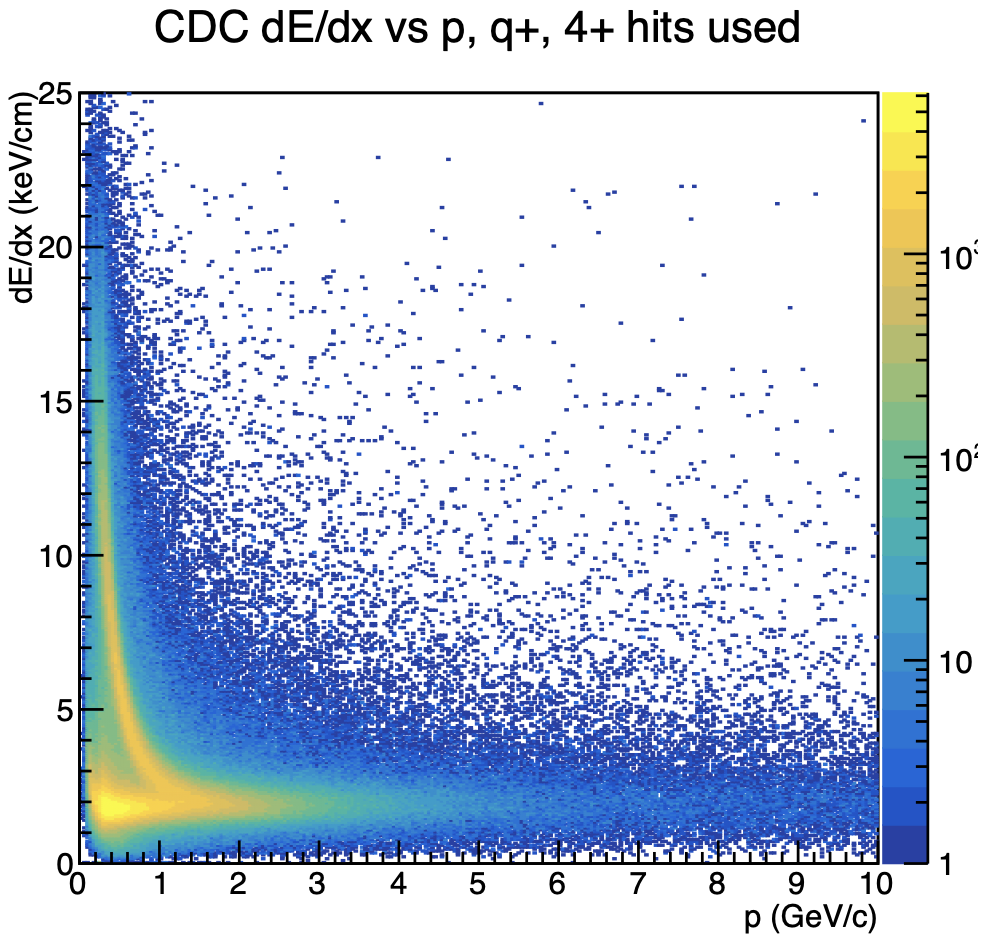}
\qquad
\includegraphics[width=.4\textwidth]{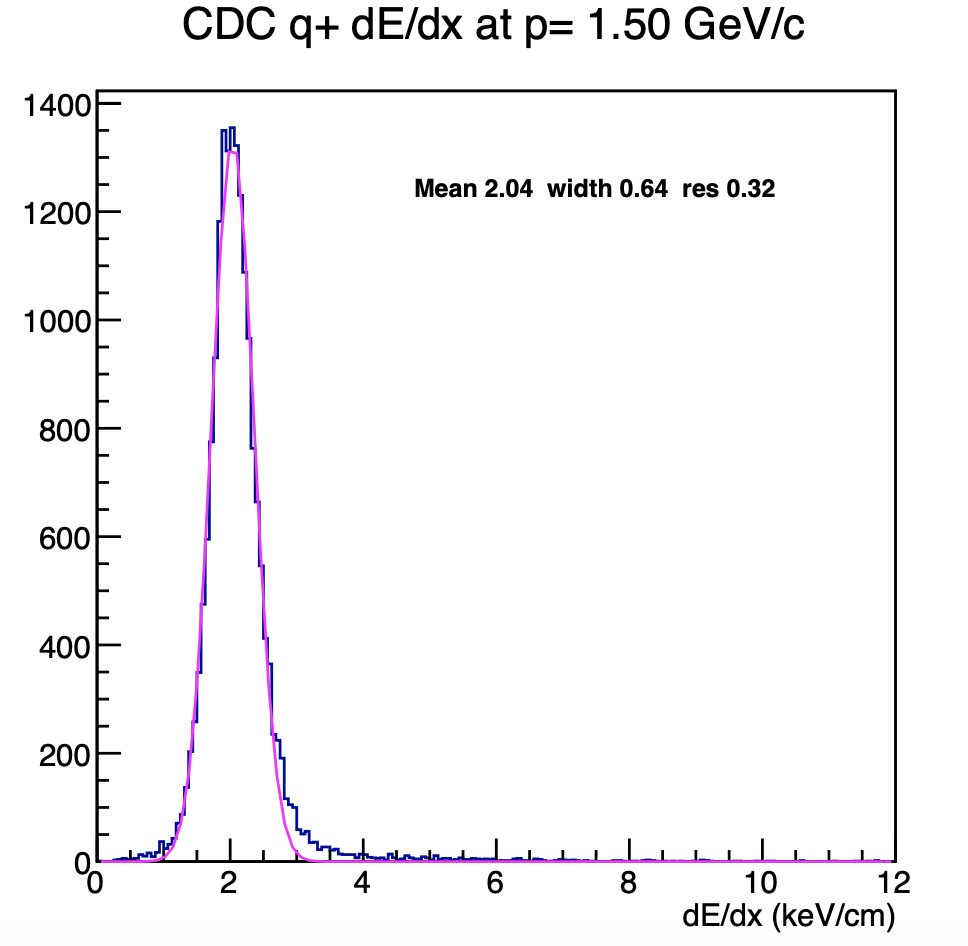}
\caption{Left: dE/dx vs p for positive charged tracks. Right: Projection of dE/dx at p = 1.50 GeV/c (blue) with Gaussian fit (magenta). The nominal peak position is 2.02 keV/cm. }
\label{fig:dedx}
\end{figure}

\begin{figure}[H]
    \centering
    \includegraphics[width=0.9\textwidth]{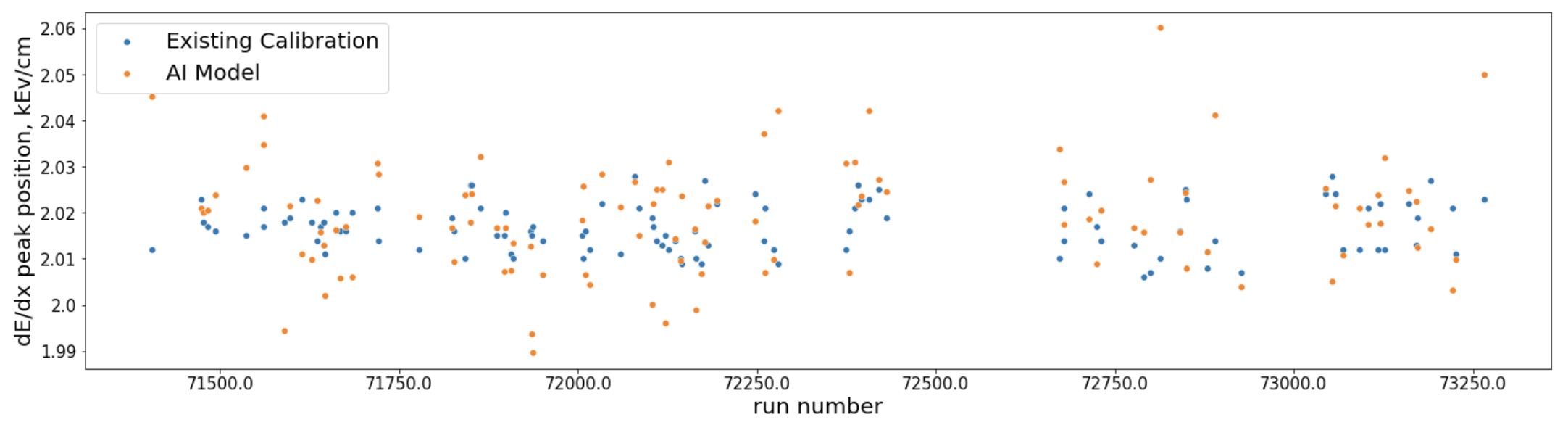}
    \caption{Mean dE/dx position with p = 1.5 GeV/c from the 2020 run period using conventional calibration (blue) and AI generated calibration constants (orange). The nominal value is set to be 2.02 keV/cm.}
    \label{fig:peak_position}
\end{figure}

Here we attempt to look at the physics data in order to see how the CDC performs using AI generated calibration constants. For the CDC, dE/dx as a function of momentum is continuously monitored. If the peak position of dE/dx at momenta equal to 1.5 GeV is far from the desired position of 2.02 keV/cm, adjustments can be made. An example of a typical plot seen during an experiment is shown in Figure \ref{fig:dedx}. As a first look at the data using our AI predicted GCFs, we can observe how this peak position would change compared to the conventionally calibrated data. This is shown in Figure \ref{fig:peak_position}. The AI generated peak positions are in general agreement with those obtained from the conventionally calibrated data.

\begin{figure}[htpb]
    \centering
    \includegraphics[width=0.7\textwidth]{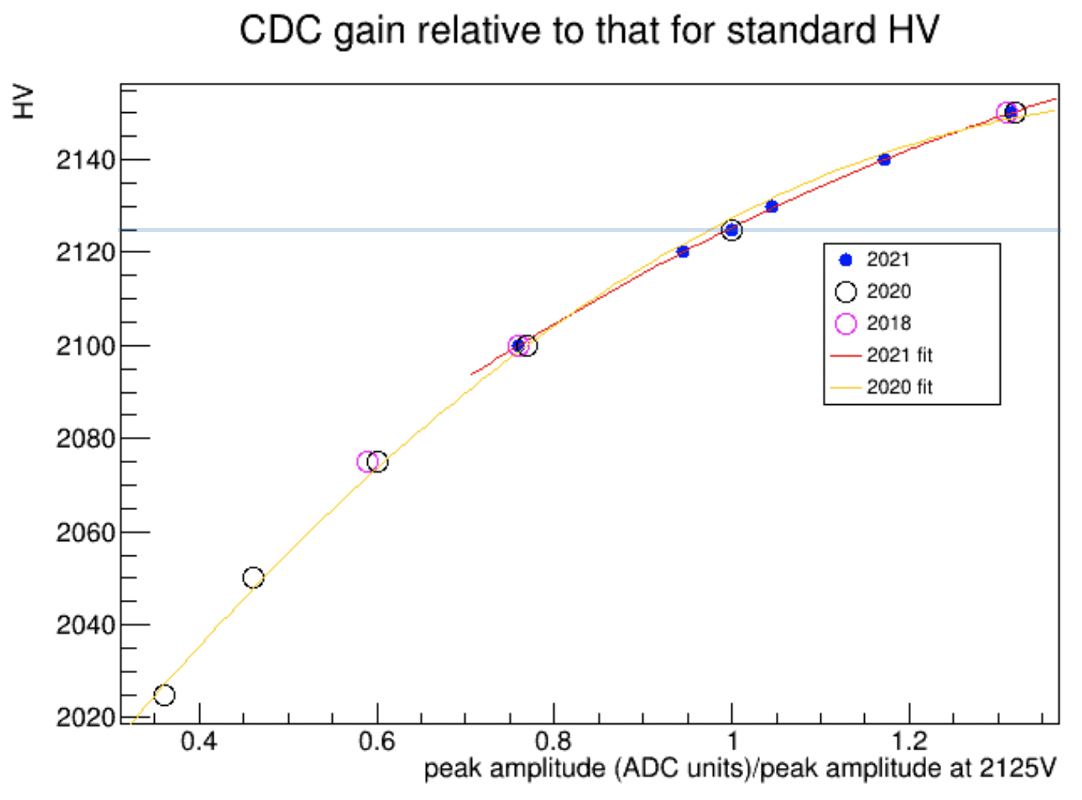}
    \caption{CDC operating voltage setting as a function of chamber gain. The data is obtained during HV scans during the 2018, 2020, and 2021 run periods. The blue horizontal line indicates the nominal CDC operating voltage. The recommendation is obtained from the fit to the data.}
    \label{fig:HV_gain}
\end{figure}

\begin{figure}[htpb]
    \centering
    \includegraphics[width=0.8\textwidth]{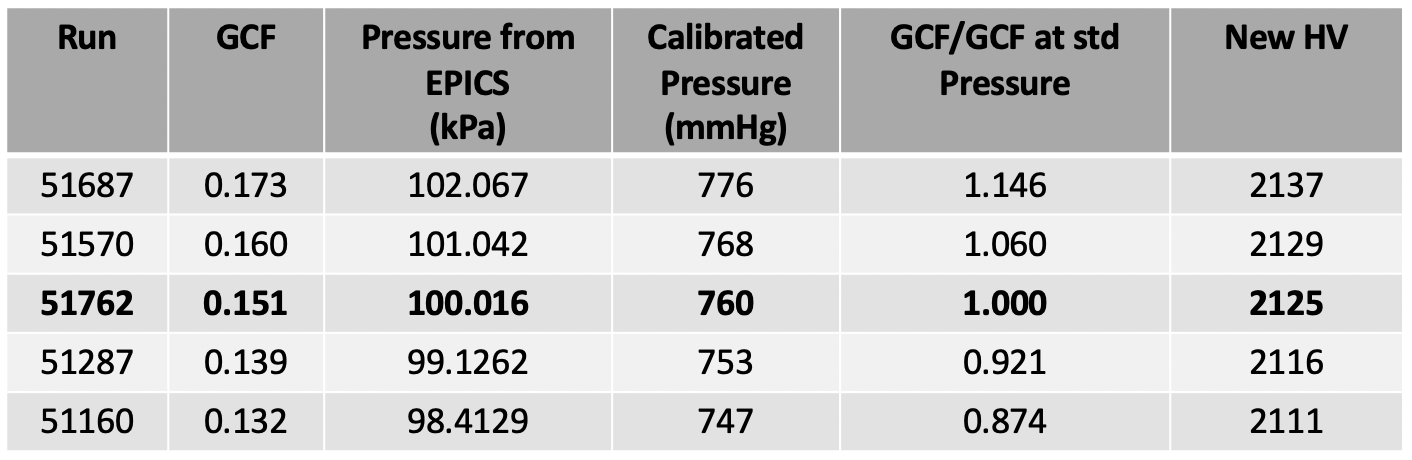}
    \caption{New operating HV values based on runs with GCFs that span the entire range of observed values. The columns from left to right are: the run number, GCF using conventional calibration, the corresponding value of the atmospheric pressure as logged by the EPICS system, the calibrated pressure value, the ratio of the GCF to that at standard pressure, and the new HV values extracted from a fit to the HV as a function of the chamber gain. }
    \label{fig:new_hv}
\end{figure}

 Figure \ref{fig:HV_gain} shows the HV setting as a function of the chamber gain obtained during HV scans for the 2018, 2020, and 2021 run periods. Once the AI has made a prediction about the GCF, the associated HV recommendation is obtained from the fit to the HV as a function of the ratio of peak amplitudes. Historically, the CDC operating voltage has always been set to be 2125 V, except during a high voltage scan. High voltage scans typically take place at the beginning of a new run period and are often the only time data is recorded at a different operating voltage. It is likely that the AI would recommend a voltage setting different from the nominal value, so investigating the effects of a different operating voltage is necessary. It is important to set the HV such that the current draw does not exceed the trip limit under normal operations. To investigate the potential range in HV values we could see, we used data obtained during previous high voltage scans to establish a relationship between the HV and chamber gain. The results are summarized in Figure \ref{fig:new_hv}. Based on the Fall 2018 run period, we do not anticipate the AI recommending voltages outside $\pm$ 20 V from the standard operating voltage.

\begin{figure}[H]
\centering
\includegraphics[width=.5\textwidth]{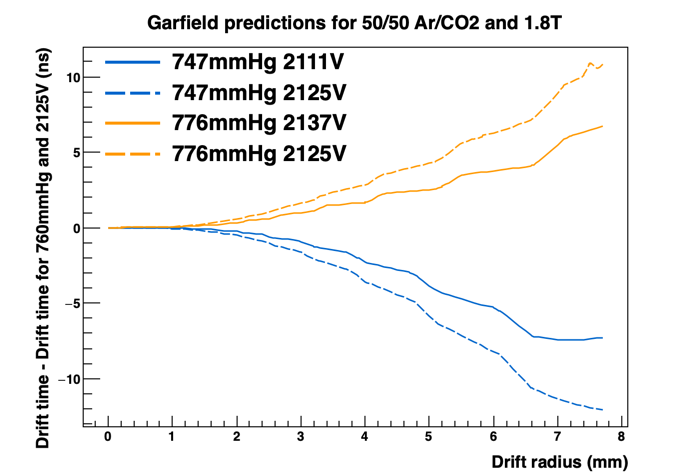}
\caption{Change in \textsc{Garfield} predicted drift times from standard atmospheric pressure (760 mmHg) and nominal operating voltage (2125 V) as a function of drift radius. The dashed curves correspond to the nominal operating voltage while the solid curves correspond to an altered HV setting.}
\label{fig:drift_times}
\end{figure}

In addition, we used GARFIELD to simulate the drift times for a 50:50 Ar:CO2 gas mixture, 1.8T magnetic field and
the values of HV that would be necessary for consistent chamber gain at extreme high and low pressure values.
The difference between these drift times and those simulated for the nominal HV and standard pressure are shown in Figure \ref{fig:drift_times}.
The differences between the drift times are small, and the drift times predicted for the extreme pressure values are closer to those
predicted for nominal HV and standard pressure when using the new HV values than when using the nominal HV.

\section{Conclusions and Outlook}
At present, we are able to predict the existing Gain Correction Factors using input features that are readily available via the EPICS system during data taking. We have established operating limits for the CDC high voltage setting in order to set appropriate constraints for the model predictions. Model development to predict the time to distance calibration constants is currently in progress. We hope to extend this system to other detector systems located at Jefferson Laboratory. 

\acknowledgments
Jefferson Science Associates, LLC operates Thomas Jefferson National Accelerator Facility for the United States Department of Energy under U.S. DOE Contract No. DE-AC05-06OR23177. This work is supported by DOE grant LAB-20-2261. The Carnegie Mellon Group is supported by the U.S. Department of Energy, Office of Science, Office of Nuclear Physics, DOE Grant No. DE-FG02-87ER40315

\end{document}